\documentclass{article}
\usepackage{natbib}
\usepackage[title]{appendix}
\usepackage{graphicx} 
\usepackage{float}
\usepackage{hyperref}
\hypersetup{colorlinks,
            linktoc=page,
            urlcolor=red,
            linkcolor=red,
            citecolor=blue}

\usepackage{amsfonts}
\usepackage{amsmath}
\usepackage{amssymb}
\usepackage{setspace}
\usepackage{authblk}
\usepackage{setspace} 
\usepackage{etoolbox}

\AtBeginEnvironment{quote}{\par\singlespacing\small}

    \title{From Protoscience to Epistemic Monoculture: How Benchmarking Set the Stage for the Deep Learning Revolution}
\author[1]{Bernard J. Koch\thanks{bernard.koch@kellogg.northwestern.edu}}
\author[2]{David Peterson}
\affil[1]{Center for Science of Science and Innovation,
Northwestern Kellogg School of Management}
\affil[4]{Purdue University Department of Sociology}
\date{April 2024}

\begin{document}
\maketitle
\begin{abstract}
 Over the past decade, AI research has focused heavily on building ever-larger deep learning models. This approach has simultaneously unlocked incredible achievements in science and technology, and hindered AI from overcoming long-standing limitations with respect to explainability, ethical harms, and environmental efficiency. Drawing on qualitative interviews and computational analyses, our three-part history of AI research traces the creation of this ``epistemic monoculture" back to a radical reconceptualization of scientific progress that began in the 1980s. In the first era of AI research (1950s-late 1980s), researchers and patrons approached AI as a ``basic" science that would advance through autonomous exploration and organic assessments of progress (e.g., peer-review, theoretical consensus). The failure of this approach led to a retrenchment of funding in the 1980s. Amid this ``AI Winter," an intervention by the U.S. government reoriented the field towards measurable progress on tasks of military and commercial interest. A new evaluation system called ``benchmarking" provided an objective way to quantify progress on tasks by focusing exclusively on increasing predictive accuracy on example datasets. Distilling science down to verifiable metrics clarified the roles of scientists, allowed the field to rapidly integrate talent, and provided clear signals of significance and progress. But history has also revealed a tradeoff to this streamlined approach to science: the consolidation around external interests and inherent conservatism of benchmarking has disincentivized exploration beyond scaling monoculture. In the discussion, we explain how AI's monoculture offers a compelling challenge to the belief that basic, exploration-driven research is needed for scientific progress. Implications for the spread of AI monoculture to other sciences in the era of generative AI are also discussed.
\end{abstract}
\section{Introduction}

Deep learning-based artificial intelligence (AI) has emerged as a defining technology of the 21st-century. This technology underpins many of today's most impressive advances in consumer, medical, scientific and even cultural technologies. It drives the internet's search and recommendation systems, tumor detectors that outperform pathologists, and algorithms that pass prestige AI challenges like Go or the Turing test with ease. As we write this, generative deep learning models are reshaping the creation of the art and poetry, writing emails, solving mathematical theorems, and calling the future of computer programming into question. 

As remarkable as these successes are, the speed at which the field of AI research (AIR) has produced them over the past decade is equally impressive. AIR's productivity is sociologically provocative because the field's approach to research differs radically from other scientific fields. Where other fields maintain loose divisions between autonomous, ``basic" research and ``applied" or ``task-driven" research (e.g., immunology v. vaccine development), the interests of academia and industry have long been deeply entangled within AIR. And where other fields assess scientific progress and significance ``organically" through mechanisms like peer review, mathematical theory-building, and citation, AIR relies mostly on the formal demonstrations of empirical performance that are common in task-driven sciences (e.g., clinical trials).

The separation of basic science from applied research in the United States emerged after World War II as an implicit contract between funders and scientists. In this agreement, scientists would be granted the autonomy to set their own agenda and evaluation mechanisms in exchange for technological products and/or policy recommendations viewed as vital for economic prosperity and military dominance \citep{bush_science_2021}. This division of rights and responsibilities allowed modern sciences to develop into largely independent fields with unique ``epistemic cultures" that could conduct research, organize, and conceptualize progress as they see fit \citep{cetina_epistemic_1999}. Although scholars in the social studies of science have complicated this neat division between ``science" and various non-scientific political, governmental, cultural, and economic forces \citep{gieryn_cultural_1999,jasanoff_containing_2009,latour_science_1987}, scientific autonomy remains a powerful rhetorical ideal. To meet scientists' side of the contract, task-driven, applied domains emerged within many fields with the aim of translating basic theoretical insights into practical technologies. 

Organizational autonomy is thought to be necessary because breakthroughs in basic research are  unpredictable and their ramifications impossible to forecast. Scientists thus need the ability to pursue intellectual interests even when possible applications are unclear. This intellectual freedom is protected by market-like, self-organizing evaluation systems such as peer review and citation. These qualitative, decentralized systems allow peer scientists to  consider a matrix of different epistemic values (e.g., accuracy, parsimony, theoretical significance, explainability) when assessing the significance of contributions. Peer review is a  deliberative process, and the elevation of high-impact works through citation is an evolving, coalescent one. The plodding speed of these evaluation mechanisms allows understandings of progress and significance to emerge \textit{organically} over time \citep{hull_science_2001,polanyi_republic_1962}.

In contrast, applied research suggests different organizational structures and evaluation mechanisms. Instead of freedom for exploration, applied science requires centralized decision-making (within or beyond the field) that can bound scientists' solution space by signaling which outcomes and epistemic values they should seek to optimize. In place of organic evaluation mechanisms like peer review and citation, task-driven science militates formal, typically quantitative evaluative mechanisms that prioritize one or a few key epistemic values (e.g., accuracy, safety). Although these metrics capture a much narrower vision of scientific progress, they have the advantage of providing immediate, explicit, and easily-interpretable feedback of progress along these dimensions.

In this article, we trace the emergence of AIR as an important and intriguing challenge to the narrative that autonomous research is necessary for progress in modern science. Through a mixed methods historical account of the development of the field through 2021, we show that AIR's most dramatic progress came after, and largely because, it ceded autonomy to external actors who set the field's agenda and established the sorts of narrow evaluation systems typical of translational science.

While AIR's productivity speaks to the strengths of this strategy, our analysis also highlights how ceding autonomy and emphasizing formal evaluation created an epistemic ``monoculture" focused only on single approach to AI: building ever-larger deep learning models. We use the term monoculture because, as we show quantitatively, the development of deep learning has come at the expense of other research trajectories in the field. In agriculture, cultivating a single crop can be incredibly productive at the expense of biodiversity. Similarly, while deep learning excels in certain dimensions, it also has significant epistemic, ethical, and environmental tradeoffs compared to other approaches that were being explored in the 2000s.\footnote{As we discuss later in the paper, the strategy of building ever-larger deep learning models is called ``scaling". Larger models are increasingly uninterpretable and require more and more electricity to train. Larger models also require more data. Cleaning internet-scale datasets of social biases can be intractable. Moreover, curating internet-scale datasets often requires privacy and intellectual property violations.} Nevertheless, its impressive accomplishments provide a strong counterpoint to the idea that undirected, autonomous research is necessary for progress in science.

Drawing on interviews with researchers across academia, industry, and government; policy makers; and tech executives; as well as archival research and computational analysis, we chronicle the emergence of deep learning monoculture from AI's formal inception as a field in 1956 to the beginning of the Generative AI era in 2021. The empirical part of the paper is divided into three sections corresponding roughly to historical periods. Each of these periods is further divided into sections discussing the epistemic, organizational, and technological aspects of the era to illustrate how these dimensions of science became increasingly aligned to set the stage for deep learning to take over the field in the 2010s.

In the first section, The Era of Symbolic AI (1955-1987), we highlight the epistemic fecundity at the origins of the field. During this period, AI researchers had a multitude of theories and research agendas for creating machines that could think. Government funders, concerned with not falling behind in the global race for technological superiority, lavished the field with huge grants and little oversight. Yet, the philosophical complexity at the heart of AI as a project led to an unproductive tribalism between competing camps. As the individual projects failed to live up to their early hype, excitement gave way to disappointment leading to the retrenchment of the second "AI Winter," in which interest and funding waned \citep{mitchell_artificial_2019}.

In the second section, The Benchmarking Era (1987-2011), we highlight how frustrations from funders led efforts to explicitly organize the field through the introduction of benchmarking, a novel formal evaluation system that presented a true alternative to the model of basic science. Benchmarking allowed AI research to elide theoretical, organization, and material issues that stymied the field in the previous decades by narrowing the epistemic vision of the field to focus only on predictive accuracy on applied tasks, and not other epistemic values prized in basic science (e.g., explainability, theoretical consistency, parsimony, efficiency). Reorienting the field toward more measurable tasks led to an influx of non-symbolic machine learning approaches grounded in probability and statistics. 

In the third section, The Deep Learning Era (2012-2021), we focus on how significant developments in both computational technology and the tech sector of the economy produced the groundwork for deep learning to become the dominant approach in AI. We argue that deep learning took over the field not because of any major intellectual advancement, but because it uniquely benefited from access to large datasets and advances in computing technology. By heavily investing in the scaling of both compute and data, large tech firms began to supplant government as the most significant external actors directing the field. Differing strategic orientations of industry compared to academia have entrenched deep learning monoculture.
 
In the discussion, we suggest that deep learning monoculture is \textit{contagious}: when scientific fields adopt deep learning for epistemic purposes, it becomes hard not to adopt some of the organizational and technological orientations of AIR's monoculture. Deep learning increasingly serves as the foundation for methodological advancement in science. However, the opacity and unexplainability of deep learning models will increase the importance of empirical benchmarking in science. By design, benchmarking canalizes science to focus on a small set of epistemic values that can be easily quantified. Moreover, because deep learning requires scale, academia will become increasingly reliant on the capacities and interests of industry to build foundation models for scientific research. The spread of monoculture via deep learning may thus alter the 20th-century model of science. Although our account is fundamentally historical, we the organizational and epistemic implications of generative AI are discussed in the conclusion.

\section{BACKGROUND: PROGRESS AND AUTONOMY IN SCIENCE}

What is scientific progress? Although there is unanimity in the belief that science makes progress, what constitutes scientific progress remains hotly debated across philosophy, science studies, and the sciences themselves. \citet{cole_making_1992}, one of the most systematic efforts in sociology to define progress, argued that sciences can be divided into research "frontiers," in which many claims compete for attention, and the research "core," which is composed of the settled and widely agreed upon "facts" constituting progress. 

Of course, this simply begs a second question; How does a claim move from the frontier to the core? Cole is noncommittal and philosophers of science and science studies scholars have been divided on this point. Early work focused on the conceptual component of theories. That is, the success of theories were taken as evidence of, in the weak version, their usefulness, and, in the strong version, their fundamental correctness. Some philosophers of science are still essentially concerned with the conceptual content of theories \citep{kitcher_advancement_1995}. Early STS was concerned with breaking the link between theory adoption and “correctness,” choosing to view scientific success through the same lens as other forms of cultural change \citep{bloor_knowledge_1991}. Scholars in this tradition have focused on the roles of technological change and social networks in achieving consensus \citep{latour_laboratory_2013}. 

More recently, scholars have argued that scientific consensus is a byproduct of advances in technology \citep{collins_meaning_1998,peterson_all_2015,rheinberger_toward_1997}. Under this view, advances in the material cultures of scientific labs extend the horizon of possibility for researchers. Novel and robust technologies that offer researchers new abilities spread through fields as other labs attempt to maintain pace. In this way, the evolution of technologies establishes a form of practical consensus. This account is largely harmonious with actor-network theorists who have grounded accounts of scientific progress in extensions across diverse social, technological, and cultural networks (e.g., \citet{latour_science_1987}). Although it is easy to caricature this tradition as supporting a “Machiavellian” sociology of science in which scientific success is indistinguishable from power \citep{fujimura_crafting_1992}, an essential insight from this tradition is that the \textit{conceptual}, the \textit{technological}, and the \textit{organizational} dimensions of scientific progress are not isolated. The categories bleed into, sometime reinforcing, sometimes undercutting, each other. Bringing this back to Cole's initial argument that claims move from the frontier to the core through an evaluation process, we suggest that understanding scientific evaluation requires an examination across all three of these categories. 

\subsection{THE ORGANIZATIONS, EPISTEMOLOGIES, AND TECHNOLOGIES OF SCIENTIFIC EVALUATION
}
Cole's argument that claims move from the frontier to the core through an evaluation process may seem a bit odd considering claims at the frontier have already surmounted what is typically thought of as the central, gate-keeping evaluation mechanism of science: peer review. If consensus is the product of simply being correct, and correctness could be immediately ascertained through peer review, we might expect there to be no gap. Yet, in fact, the peer-review process can often take months or years. Claims may be more fragile or less significant than initially thought. Over time, it can become clear that a new innovation is less novel or useful than originally claimed. 

Because of this, evaluation in science is a longer and more complex process than proponents of peer review suggest. We label this ideal-typical form of evaluation\textit{ organic}.\footnote{The language of “organic” and “formal” in the domain of scientific evaluation first occurs in \citet{peterson_self-correction_2021} to describe mechanisms of self-correction. We expand the usage here to apply to consensus-building more broadly and we argue that these forms of evaluation different in important ways across epistemic, organizational, and technological dimensions. } \textit{Organizationally}, organic evaluation implies an autonomous, self-organizing field \citep{hull_science_2001,polanyi_tacit_1962}. \textit{Epistemologically}, it evaluates claims within a matrix of multiple theories and epistemic values. \textit{Technologically}, organic evaluation is characterized by the slow, uneven. spread of technological innovations across the field’s social networks. 

Three things should be clear from the above description. First, the process is decentralized. After the bottleneck of peer review in which a small group of 3-5 core set members evaluate whether a paper deserves to be published, the field as a whole decides the worth of a claim by choosing to integrate into their own practice or ignore it. Second, there is no single or even small number of values that is used. Claims are evaluated on the basis of novelty, significance, robustness, cost, efficiency, and many other values./footnote{Although \citet{lamont_how_2009} on grant review panels focuses mainly on social sciences and humanities, it is a useful touchstone for the many competing values that go into disciplinary evaluations.} No quantitative measure of “best” is appropriate in judgment domains where competing values need to be weighed. Third, because it involves changes in scientific practice, organic evaluation, as a process, involves temporal extension of an unpredictable duration as researchers attempt to import new technologies and techniques and try to follow the pathways suggested by a claim. Initial excitement can founder as members of a field come to understand a claim is weaker, more limited, or less useful than thought. And even successful adoption may take years depending upon the technical challenges involved. 

However, there are many situations where people want evaluation to provide unambiguous, immediately actionable information based upon specific uses. While a microbiologist may pursue theoretically-driven research about the behavior of a class of bacterium which may be evaluated in a qualitative process extending over months and years, when that research becomes the basis for clinical intervention we demand evaluation to be clear and timely. 

Under these conditions, \textit{formal evaluation} methods are often adopted. Formal evaluation differs from organic along each dimension. \textit{Organizationally}, formal evaluation requires centralization. \textit{Epistemologically}, the matrix of values collapses upon a single or small set of values to provide clear feedback along a dimension of interest. \textit{Technologically}, a high level of standardization is required to reduce the ambiguity in evaluation that \citet{collins_changing_1992} has famously referred as the “experimenters’ regress.” 

The division between organic and formal evaluation regimes is related to long-standing discussions about basic v. applied or understanding- v. task-oriented science. Scientific projects that pursue basic research for purposes of understanding are generally evaluated organically, while those focused on specific applications benefit from more formal approaches. Significantly, however, these divisions do not characterize entire sciences. It is not like ecology favors organic evaluation while economics favors formal. Rather, most fields are engaged in both projects and, thus, utilize evaluation techniques across the spectrum depending upon whether the project is pursuing autonomously chosen basic research or projects that are directed with reference to specific outcomes or goals. 

Yet, there is an inherent tension between these evaluation regimes. On one hand, because of the lack of clear metrics in organic evaluation, it can be unclear whether progress is actually being made or if such progress is actually useful. And the extended time frame needed to resolve those questions can be frustrating for outsiders who rely on that information, and funders who want to evaluate their return on investment. On the other hand, formal evaluations can provide clear metrics of progress. But by embedding specific values into the evaluation mechanism, it strips fields of the autonomy that allows them to qualitatively evaluate their own products. Thus, debates over evaluation become proxies for debates about the autonomy of fields. 

\subsection{EVALUATION AND AUTONOMY}
Although their sociologies of science offer stark contrasts with each other, Merton and Bourdieu shared the belief that an essential and distinguishing feature of scientific communities is the fact that scientists primarily produce research for an audience of fellow scientists. As \citet{bourdieu_science_2004} expresses most emphatically, “The first and probably most fundamental of the distinctive properties of the scientific field is, as we have seen, its (more or less total) closure, which means that each researcher tends to have no other audience than the researchers most capable of listening to him” (69).

Although one might reasonably argue that science operates best when allowed to set its own agenda and evaluate its own products, external “guiding” and “shaping” are an important feature of science funding and policy. Although much evaluation occurs within one’s field (e.g., peer review and hiring), many forms of evaluation are conducted by external audiences who must decide for themselves the value of a particular field, or scholar, or claim.

Moreover, the post-WWII commitment to science involved the expectation that the funding and autonomy useful for the development of systematic theory would \textit{eventually} produce tangible technological, military, economic, and social benefits. Thus, the practical application of science was always part of the deal and this translation of basic science to real world settings always involved some ceding of authority.

That said, the space where evaluation meets external audiences is an inherently fraught zone. Rather than speaking to colleagues and other members of the “core-set” of a field \citep{collins_place_1981}, interactions with funders, industry, the government, or the public sets up a risky interaction for the scientist as the “pure” motives of the scientist get undermined and swayed by the idols of the market, theater, and tribe. This is a reason why \citep{abbott_status_1981} argues that the highest status members of professions are those best able to insulate themselves from those who utilize the expert’s expertise. Rather than be tainted by association with values and motives orthogonal or anathema to their field’s, they are able to withdraw into a space of “professional purity.”

Because they tend to span basic and applied projects and, thus, engage both organic and formal evaluation mechanisms, most sciences have domains where they are largely autonomous and self-directing, as well as places with diminished autonomy. In early accounts, these basic research projects produced the knowledge that would be adapted by applied scientists and engineers as exportable technologies. Subsequent work in the social studies of technology have complicated the picture, suggesting the causal arrow works in both directions. Yet, throughout this literature, fundamental, basic, theory-driven research is always considered to be an essential part of scientific progress (if not \textit{the} essential part). 

Yet, this need not be the case. As we will show, modern machine largely emerged out of a regime of task-driven, formal evaluation. In the following sections, we detail the historical emergence and structuring of the field of AIR through three periods. In the first period, the largely autonomous field pursued basic research, but ultimately succumbed to an “AI Winter” because the mechanisms of unstructured evaluation proved ineffective at achieving consensus. In the second, DARPA, the major funder in the field, reoriented their funding from large, no-strings-attached, ``blue sky" grants, toward funding narrowly defined, well-specified tasks with clear metrics of progress. These highly structured, quantitative evaluations, called benchmarks, allowed the field to meaningfully chart progress and compare competing approaches. Yet, advances in data access and compute power eventually led to the domination of deep learning, a machine learning approach uniquely positioned to take advantage of those improvements. The success of deep learning on quantitative benchmarks soon led to the crowding out of nearly all competing approaches. The third section details how machine learning has become an \textit{epistemic monoculture}, a knowledge-producing community organized around a handful of organizations, a constrained research trajectory, and a single epistemic value.

Ultimately, we argue that AIR represents the vanguard of a new scientific movement that views task-driven science not as a derivative of basic research, but as a unique mode of pursuing progress with both dazzling advantages and significant risks.

\section{METHODS}

\subsection{INTERVIEWS}
Our analyses draw primarily on 45 in-depth, open-ended interviews with individuals spanning the history and breadth of AIR. Many of the most influential figures in the first era of our study (1956-1986) are now deceased, so we spoke with academics who attained PhDs in their symbolic AI groups and departments before the transition to statistical learning in the late 1980s. We supplement these interviews with archival sources (see below).

Our portrayal of the second period (1986-2010) is built on conversations with individuals in government, academia, and industry who were pivotal to the transition from symbolic AI to machine learning. These include program managers and scientists at the Defense Advanced Research Projects Agency (DARPA) and the National Institute of Standards and Technology (NIST) who were personally responsible for introducing formal evaluation mechanisms to the field, as well as leading industry researchers who either invented or introduced key statistical learning algorithms (e.g., hidden markov models, convolutional neural networks, random forests) at Bell Labs and BBN Technologies. We also spoke to a range of academics, including the first organizers and editors of the Conference on Machine Learning (now ICML) and Journal of Machine Learning in the 1980s.

For the third era (2011-2021), we had broad access to individuals across government, industry and academia with a wide range of perspectives on deep learning. Highlights include PhD researchers and research managers at Google Deepmind, Microsoft Research, Apple Research, and Spotify Tech Research. We also spoke to experts on benchmarking in the deep learning and generative AI eras at NYU, Google, Stanford, the Santa Fe Institute, and the Allen Institute for AI. With respect to policy, we spoke with a white-house level advisor on AI at NIST and a VP in charge of research at Microsoft Nuance. Finally, we spoke to prominent AI ethicists and deep learning critics at Google Deepmind and the University of Washington. 

\subsection{ARCHIVAL RESEARCH}
To get a better sense of the key issues during the early decades of AI, we relied heavily on interviews and original documents in the Edward Feigenbaum Archives at Stanford University, editorials in AI Magazine from the 1970s-1990s (the general interest professional publication of the Association for the Advancement of Artificial Intelligence), and recordings/transcripts of talks by scientists and critics from that era. We also draw examples from publications in AI and ML at major workshops/conferences from 1960s-2020s anthologized in print, online conference proceedings, and more recently, arXiv. Lastly, we sometimes reproduce quotes from leaders in the field drawn from secondary sources.

\subsection{COMPUTATIONAL ANALYSIS}
We corroborated some of the themes that emerge from our qualitative research using 90,000 papers published between 1993 and 2018 from Microsoft Academic Graph in 14 top AI research venues: AAAI, ACL, AISTATS, COLING, CVPR, ECCV, EMNLP, ICCV, ICML, KDD, NAACL, NeurIPS, SIGIR, and WWW. We searched for keywords related to five different AI approaches in paper titles, abstracts, or topics from a labeled topic model. We used LiteRate, a method for time series analysis, to estimate the underlying rates at which new papers were published on each of these topics over time (\citet{koch_evolutionary_2020}, under review). LiteRate provides credible intervals for these estimated rates, as well as statistically significant rate shifts which are indicative of major historical events.

\section{1956 TO THE MID-80s: THE ERA OF SYMBOLIC AI}

Propelled forward by the post-war futurism of Alan Turing and John Von Neumann, AI emerged as a nascent field in the 1950s with the promise of becoming, in Herbert Simon’s (1970) phrase, “a science of the artificial.” But as Herbert Simon later recalled, “AI has had problems from the beginning. It is a new field in which people came from many different directions” \citep{simon_artificial_1995}. In the following decades, the field was unable to realize itself as either basic or task-driven science. Instead, it stagnated as a “proto-science” like philosophy, where opposing intellectual projects represented by charismatic leaders evolve in parallel, unable to build progress towards common theoretical language or empirical methods that can adjudicate debates \citep{collins_sociology_1992}. \textit{Epistemically}, the field sprouted a variety of quasi-realist tendrils inspired by the workings of the human brain or mind. However, these programs were stymied by our poor understanding of either. \textit{Technologically}, the limits of 20\textsuperscript{th}-century computers confined researchers to working on “toy” examples that were unsatisfactory to test their theories. And \textit{organizationally}, the field was unable to unify the fractured, personality-driven social networks that emerged at its inception. This divided social structure was encouraged by the exceptional autonomy AI researchers enjoyed from funders. Although it was fueled by hype around potential, the field never developed rigorous mechanisms of evaluation nor a shared agenda. The arrangement led to a series “AI Winters” in which funding dried up.

\subsection{THE EPISTEMIC TRIBALISM OF EARLY AI}

The origins of AI as an academic field are commonly traced back to The Dartmouth Summer Research Project on Artificial Intelligence that took place in 1956. The workshop’s goal, according to John McCarthy and his fellow organizers, was to convene a group of researchers who would “proceed on the basis of the conjecture that every aspect of learning or any other feature of intelligence can in principle be so precisely described that a machine can be made to simulate it” \citep{mccarthy_john_proposal_1955}. The conference was attended by several influential 20th century computer scientists, including those who would dominate AI for the next four decades: Allen Newell, Herbert Simon, John McCarthy, and Marvin Minsky. In the years following the workshop, these figures established a wide array of research programs consistent with different visions of the workshop’s mission.\footnote{
    At Carnegie Mellon, Simon and Newell cultivated a cognitivist approach to AI that focused on emulating how humans think. For example at Dartmouth, they showcased their “Logic Theorist” algorithm that proved theorems not by brute force exploration of solutions, but by using heuristics to efficiently explore possible approaches (Frants, 2003). Initially at MIT and later at Stanford, McCarthy believed that the highest form of intelligence, reasoning, proceeded through internally-consistent rules of logic. His commitment to formal logic led him to develop the LISP programming language as a way for computers to manipulate math. At MIT, Minsky was inspired by the modularity of the human brain. He argued that intelligence would emerge through the complex interactions of limited modules focused on specific problems.
    
    While they lacked the longevity of the above programs, other peripheral projects in the 1950s and 1960s were inspired by still other aspects of intelligence. At IBM, Arthur Samuel developed a checkers playing algorithm that served as the basis for reinforcement learning. At IBM and Stanford, researchers Raj Reddy and Yehoshua Bar-Hillel were respectively inspired by the new field of linguistics to create algorithms for machine translation and automated speech recognition. At Cornell, psychologist Frank Rosenblatt picked up on early work by McCullough and Pitts to advance a learning approach inspired by the human brain called “artificial neural networks.” Rosenblatt demonstrated some early successes in computer vision using the famous perceptron algorithm.
}

The unruly diversity of these projects underscored what ultimately proved to be the essential, fatal problem of the field: From Dartmouth onward, nobody could agree on what the term “artificial intelligence” actually meant. Most of the Dartmouth attendees at least agreed that the “artificial” part should be pursued through serial manipulation of symbols and concepts (Minsky, McCarthy, Simon, and Newell). But mathematical psychologists like Frank Rosenblatt argued that sub-symbolic, parallel processing by brain-like “neurons” (first suggested by McCulloch and Pitts) was a compelling alternative \citep{mcculloch_logical_1943,rosenblatt_perceptron_1958}. “Intelligence” was an even deeper quagmire. For example, McCarthy and Minsky were initially inseparable as assistant professors at MIT, but grew apart because McCarthy’s faith in logical reasoning was incongruent with Minsky’s belief that intelligence emerged from complex interactions. The rift ultimately led McCarthy to found his own logicist school at Stanford. When asked about the division between the tribes, McCarthy lamented,

\begin{quote}
I think we don’t talk to each other as much as we should. We tend to have these separate empires […] none of us has an excessive talent in understanding other people’s points of view. There’s a tendency after starting a discussion to say, ah yes, this suggests something that I want to work on, and the real desire is to get off alone and work on it \citep[pg. 133]{mccorduck_machines_2004}.
\end{quote}

Minsky, ever the provocateur, had sharper words:

\begin{quote}
McCarthy has tried to isolate individual fragments of knowledge and see if he could avoid the somewhat ad hoc packaging of them into larger chunks. If he wins it will be nice, but he can’t. I’m pretty sure of that \citep[pg. 308]{mccorduck_machines_2004}.
\end{quote}

In many sciences, shared epistemic values and experimental cultures provide a productive avenue to resolve disagreements \citep{kuhn_structure_2012}. But in AIR, there simply was no consensus on which values (e.g., brain-like realism, mind-like realism, parsimony, or mathematical formalism) were the most important ingredients of a good theory. Appealing to realism as evidence for the superiority of one's approach was particularly difficult because the mind and brain remain poorly understood (cognitive science and neuroscience were even more immature than AI). When complex problems are amenable to multiple empirical strategies, scientists sometimes prefer to distance themselves from competing research programs \citep{peterson_all_2015}. In AI, the early creative and intellectual differences evident at Dartmouth thus ossified into distinct schools more akin to pre-paradigmatic philosophical tribes led by charismatic personalities, rather than unifying into a coherent paradigm \citep{collins_sociology_1992}.

\subsection{THE TECHNOLOGICAL LIMITATIONS OF EARLY AI}

A second reason that AI was unable to coalesce around any one approach is that technological constraints prevented them from empirically testing or realizing their theories. Collins describes the advancement of machinery as the key ingredient for rapid discovery \citep{collins_why_1994}. Yet, despite significant gains over fifty years, 20\textsuperscript{th} century computers were consistently too anemic to actualize the field’s grand ambitions. This led to a focus on limited and self-contained "toy problems” which, although they held little intrinsic value, were supposed to showcase the potential for different approaches. Examples included puzzles like Simon’s Tower of Hanoi, chess scenarios with a limited number of starting moves, or Minsky’s artificial “block worlds” where virtual robots manipulated virtual objects \citep{simon_functional_1975,winograd_procedures_1971,winston_learning_1970}. 

In an op-ed in AI Magazine in 1991, Roger Schank, a central figure in natural language processing at the time, retrospectively delineated “real” AI from mere software engineering by emphasizing that AI “program[s] [are] based on a theory [or algorithm] that is likely to scale up” \citep{schank_wheres_1991}. But the tacit assumption that success on toy problems would generalize (i.e., “scale up” in Schank’s words) to more complex problems was rarely supported. Rule-based, symbolic approaches in AI have proved especially susceptible to the combinatorial explosion. That is, as research moves from the Tower of Hanoi to checkers to chess, the complexity of the task does not increase linearly. It mushrooms exponentially. 

In AI, the theoretical and material constraints were so intractable that researchers struggled to appeal to epistemic values to justify their work. Instead, hyping the success of toy problems was one of the few ways that researchers could continue to justify interest and investment in the field. Leaders were thus some of the worst perpetrators of this practice. For example, Simon mused in 1957 that an AI algorithm would be a world chess champion within ten years \citep{simon_heuristic_1958}.

Schank called hype around the capabilities of toy examples the “gee-whiz” definition of AI \citep{schank_wheres_1991}. Bar-Hillel, an early innovator in automated speech recognition, pessimistically labeled the irrational belief that toy problems would lead to artificial general intelligence, “the fallacy of the first step” \citep{dreyfus_history_2012}. In other words, simply because AI researchers build a machine that can perform some human-like action does not mean a sure path forward has been discovered. To put it more bluntly, fiery AI critic Hubert Dreyfus quoted his brother, computer scientist Stuart Dreyfus: “It’s like claiming that the first primate to climb a tree was taking a first step towards flight to the moon” \citep{citris_keynote_2014}.

Panofsky has documented how dysfunctional protosciences can court outside attention to maintain legitimacy \citep{panofsky_misbehaving_2014}. Without empirical evidence, hype around first-step fallacies became a potent form of social currency within the field. Frank Rosenblatt, the chief advocate for artificial neural networks (ANNs) in the 1950s-1960s, courted \textit{Science} and the \textit{New Yorker} with toy examples of the perceptron learning “representations” of cats from images. Under the title “Human Brains Replaced?,” \textit{Science} hyped the invention as “no ordinary mechanical mind which stores up information and regurgitates it [, the] Perceptron may eventually be able to learn, make decisions, and translate languages” \citep{crevier_ai_1993}.

Because hype was such potent currency in the field, Minsky felt threatened by Rosenblatt’s limelight, and disliked him personally because of the lack of mathematical rigor in his work \citep[p. 106]{mccorduck_machines_2004}. In 1969, he and Seymour Papert published a book, called \textit{Perceptrons}, condemning “most writing" on Perceptrons for being "without scientific value”  and providing a famous proof showing that perceptrons were incapable of solving a class of simple problems with nonlinear solutions  \citep{minsky_perceptrons_1969}. The book was so caustic that it significantly wounded Rosenblatt both personally and professionally \citep{mccorduck_machines_2004}.\footnote{The book apparently deeply affected Rosenblatt, and he died just two years later in a boating accident on his 43\textsuperscript{rd} birthday.}

\subsection{ORGANIZATIONAL AUTONOMY AND DIVISION}

Without theoretical consensus or empirical progress, fields can fail to maintain the interest of funders. Yet the Defense Advanced Research Projects Agency (DARPA), the field’s key 20\textsuperscript{th} century benefactor, granted exceptional autonomy to elite AI researchers from the 1950s through 1980s.

DARPA (then called ARPA) was founded in the aftershock of the Soviet Sputnik Launch (1958). AI was seen as a key dimension of the US’s competitiveness in the science race. Licklider, the visionary director of ARPA’s early computing initiatives from 1962-1964, held the philosophy to “Fund people, not projects!” \citep{crevier_ai_1993,salisbury_cautionary_2020}. He put his money where his mouth was starting with a gigantic, 2.2 million dollar grant to MIT in 1963  (\$20+ million in 2023), much of it flowing to Minsky \citep[pg. 65]{crevier_ai_1993}. For the next three decades, grants of two to three million dollars came annually to elites at CMU, Stanford, and MIT with limited strings attached. Edward Feigenbaum was a student of Simon and the father of expert systems, the most significant commercial application that emerged from symbolic AI. He noted, 
\begin{quote}
ARPA was focused on excellence. Now, people outside the ARPA community of investigators would say that that's an insider's view, that in fact it was highly political; it was just that there was an "in" group and an "out" group. If you were a student of Newell, or Simon, or Minsky, or Fano, you were in. That was very political \citep{feigenbaum_interview_1989} .
\end{quote}
According to Feigenbaum, it wasn’t merely that ARPA funded specific people. In the heady days of the early 1970s, those same people were in fact, setting ARPA’s agenda. Feigenbaum reflected wistfully on the days when a small group of AI researchers organized the entire field,
\begin{quote}
The ARPA PI's got together in a room for a few days and talked science, and helped the ARPA people plan their next year or two of projects […] We had Allen Newell, and Alan Perlis, and John McCarthy, Marvin Minsky, myself…. those were terrific -- 20 or 30 people; absolutely stellar, wonderful meetings \citep{feigenbaum_interview_1989}.
\end{quote}

The field evaded accountability partly because there was always a geopolitical reason to fund AI. In the 1980s, Japan’s ten-year “Fifth Generation” initiative to build supercomputers for future AI motivated DARPA to invest \$1 billion on the “National Strategic Computing Initiative” (NSCI) \citep{stefik_strategic_1985}. And, by this time, DARPA had concluded that AI’s rare accomplishments (e.g., a missile targeting system) justified their “blue-sky” granting philosophy \citep{forsythe_studying_2001}.

According to a DARPA-commissioned history of NSCI, program managers during this period did not even have the technical expertise to assess the work they were funding \citep{roland_strategic_2002}. And, because there was a “code of silence” among the grantees “not to embarrass each other” in front of the program manager, the information they used to assess progress,  
\begin{quote}
came from atmosphere, body language, perspiration, and nervousness. One could watch the speaker or watch the other PIs. It wasn’t hard to distinguish between the confident, enthusiastic presentation that held the audience and the halting and wordy meandering that lost or confused the audience. In a way, the other [grantees] were the [program manager's] best guide to what was good and what was not \citep[pg. 204]{roland_strategic_2002}.
\end{quote}

In 1971-1974 (the first “AI winter”), AI researchers across the US and UK did come under fire for failing to deliver on their large promises. But DARPA insulated the elites. Researchers were forced to frame their projects in terms of military relevance even though, to quote Feigenbaum, “what we all knew about the military could be put in a thimble” \citep{feigenbaum_interview_1989}.  When the DARPA program officers read these awkward attempts to appeal to their military patrons, they simply wrote the justifications themselves: 

\begin{quote}
…the DARPA people looked at it and said, "That's silly. We can't write that stuff. This is ridiculous. We'll have to take over that job for you." So they started to craft all the language, which they would then slap on to our proposals. We would never see that language. They would write all that stuff. The only people who ever saw it were the students who would later dig it up under Freedom of Information Act. It wasn't in the Stanford version of the proposals. It was only in the stuff that went up to the DARPA front office \citep{feigenbaum_interview_1989}.
\end{quote}
The lack of accountability in AI is important because with external accountability comes evaluation, and evaluation systems drive consensus formation.  Organized evaluation is a defining social structure of science that gives it both epistemic and social legitimacy. Yet for the first forty years of AI’s history, researchers were able to trade in hype rather than measured progress. In 1984, McCarthy made an impassioned plea for more rigorous evaluation in a AAAI presidential address \citep{mccarthy_we_1984}. But by the time Cohen and Howe wrote a manifesto on how to do rigorous evaluation in AI in 1988, it was too late. A bitter cold had descended on the field \citep{cohen_how_1988}.      

\subsection{THE AI WINTER OF THE LATE 1980s }

Without evaluation, AI researchers lost control of the hype train when “expert systems” caught fire in the late 1970s. Pioneered by Feigenbaum, expert systems were logical induction and deduction systems that could draw conclusions through if-then-reasoning from facts encoded in a knowledge database. The technology proved successful at sophisticated but constrained tasks, such as reading mass spectrometry reports or determining different types of loans a bank customer is eligible for. Investment into companies marketing expert systems for complex, ambitious problems like medical diagnosis exploded.

But according to Schank, the technology was fatally flawed because it could not naturally generalize; each knowledge relationship had to be laboriously hand-encoded by human experts and knowledge engineers \citep{schank_what_1987,schank_wheres_1991}. This created brittleness that resulted in spectacular failures, such as a humorous vignette by Forsythe about a medical system that diagnosed a cis-gender man with a disease of the female reproductive system because the relationship between sex and those diseases had not been explicitly encoded \citep{forsythe_studying_2001}. By 1984, Schank and Minsky were both warning the public that expert systems were 15 year old technology and the “intelligence” in these systems was limited, but it was too late \citep[pg. 204]{crevier_ai_1993}. Ultimately, the field collapsed beneath the limitations of expert systems, shifting market forces, and the crushing weight of expected advances that never materialized. Economically, expert systems were also increasingly outcompeted by cheaper personal computers.

The confluence of these factors led to a cataclysmic “AI winter” in both academia and industry. AI start-ups shuttered and the money dried up. As it became clearer that the weaknesses of symbolic AI represented significant technological challenges with no clear path to resolution, the no-strings attached grants and plentiful academic positions also disappeared. Researchers fled to industry. According to John Makhoul, a scientist working in automated speech recognition at the time, "'AI' became a dirty word, because of overhyping."

\section{1980s TO 2012: THE BENCHMARKING ERA}

The vacuum created by GOFAI did not set the stage for the natural emergence of a new basic science approach to general AI. Instead, the hard reset allowed DARPA to experiment with a radical new formal evaluation system called “benchmarking” for applied problems of obvious commercial and military utility; problems like machine translation, information retrieval, and visual object recognition. Task-driven scientific communities generally draw from a foundation of knowledge built by a basic science wing of their field. What made this experiment revolutionary is that benchmarking presented an alternative vision of scientific progress. Rather than a slow evaluation process occurring within a matrix of competing values, benchmarking valorized a single epistemic value, predictive accuracy. This allowed AI communities to elide the intractable theoretical and material quagmires that had plagued GOFAI, and make progress through a suite of statistical learning approaches now collectively called “machine learning.” Because benchmarking required ceding autonomy to external actors and the displacement of organic evaluation, we view the spread of this mono-value optimization system as one of the primary seeds of monoculture within the field.

\subsection{THE ORGANIZATION OF THE FIELD THROUGH BENCHMARKING}

As GOFAI imploded, one of the applied problems that re-piqued DARPA’s interest was automatic speech recognition (ASR). ASR was perceived to be a risky investment at DARPA. In what should now be a familiar story, ASR research had been weakly funded since a high-profile report lambasted the field, comparing it's ambitions to “schemes for turning water into gasoline, extracting gold from the sea, curing cancer, or going to the moon” during the first AI winter \citep{pierce_whither_1969}.\footnote{Mark Liberman shared a detailed history of automated speech recognition with us that illustrated this point. In the 1950s, there was exuberant optimism about how theories of phonology and grammar from the new science of linguistics would crack machine translation and speech recognition. But as evidenced by the Schank quote above, the actual performance of rule-based linguistic methods was anemic, and optimism began to sour in the 1960s.

In 1966, John Pierce, a respected executive at Bell Labs who supervised the invention of the transistor, wrote a polite but pessimistic report about the future of machine translation for the National Academies of Sciences (Liberman, 2015). In 1969, he wrote a significantly more caustic letter to the Journal of the Acoustical Society of America slamming ASR:

\begin{quote}
We are safe in asserting that speech recognition is attractive to money. The attraction is perhaps similar to the attraction of schemes for turning water into gasoline, extracting gold from the sea, curing cancer, or going to the moon. One doesn’t attract thoughtlessly given dollars by means of schemes for cutting the cost of soap by 10\%. To sell suckers, one uses deceit and offers glamor \citep{pierce_whither_1969}.
\end{quote}

Aside from one last play at ASR using GOFAI from 1972-1975, Pierce’s criticism largely killed interest in funding computational linguistics from 1970-1986.
}

\begin{figure}
    \centering
    \includegraphics[width=\linewidth]{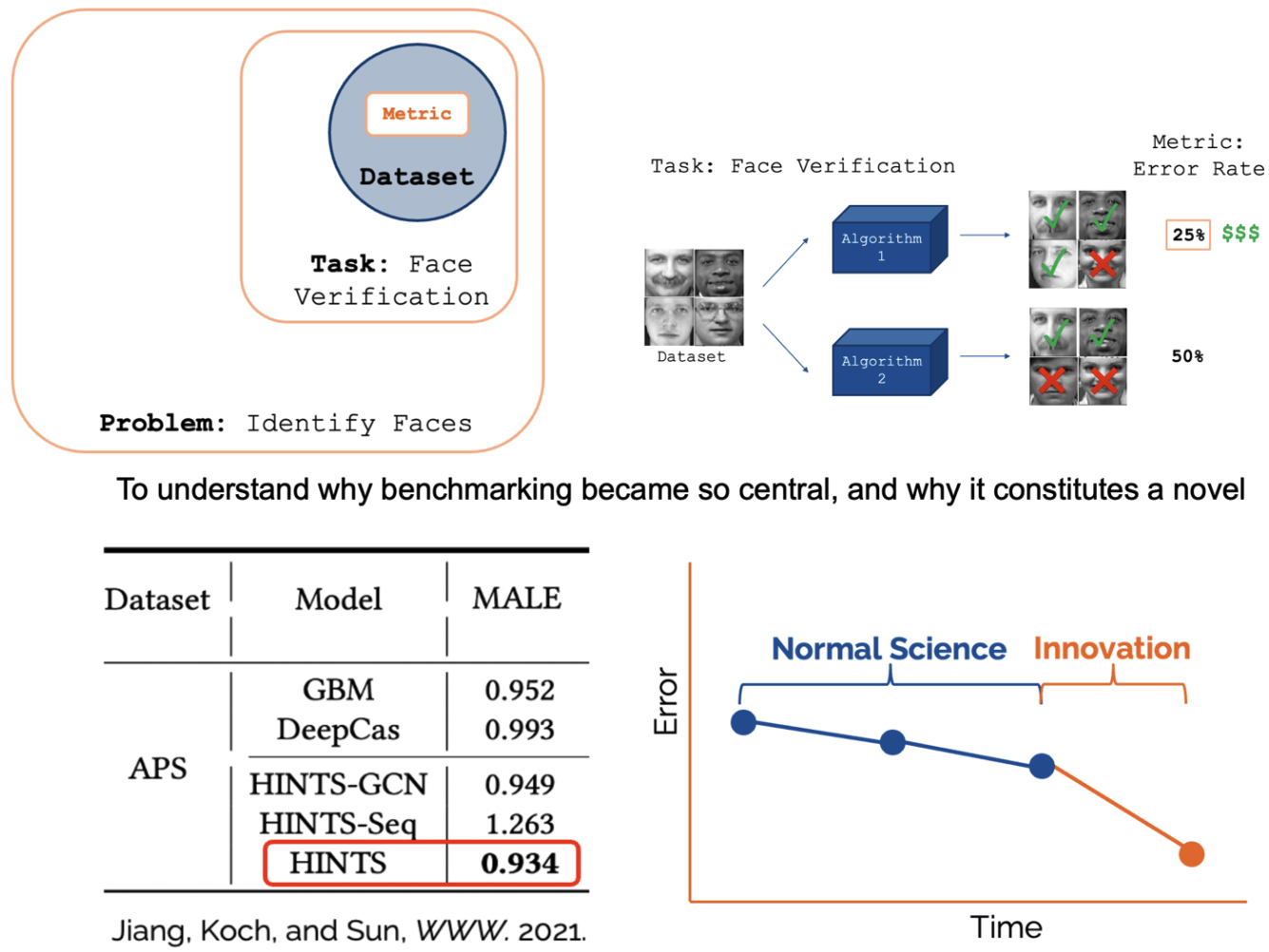}
    \caption{\textbf{The process of benchmarking}. \textbf{Top Left:} A “benchmark” consists of a task that is part of a larger a problem, a dataset that is representative for that task, and a metric (usually some version of accuracy) that scientists must build an algorithm to maximize. \textbf{Top Right: }Scientists then set their algorithms to compete on the benchmark. The algorithm with lowest error (highest accuracy) wins the grant. \textbf{Bottom Left: }A typical benchmarking table that appears in a machine learning paper. Authors bold their algorithm’s scores to highlight that they achieved state of the art accuracy/error scores. \textbf{Bottom Right:} Hypothetical benchmarking curve for a task community over time. Gradual lowering of the state-of-the-art error score is “normal science” in Kuhnian terms. A large jump in the state-of-the-art suggests a significant innovation.}
    \label{fig:benchmarking}
\end{figure}

With only a small budget in 1985, program manager Allen Sears proposed a new evaluation system that would allow DARPA to immediately quantify return-on-investment. DARPA would hold a competition in which researchers competed to build an algorithm that could achieve the highest accuracy score on a previously unseen dataset (Fig. \ref{fig:benchmarking}). In ASR, this meant building an algorithm that could transcribe the highest number of words from an independent recording correctly. The winner of the competition would get the grant, but all researchers were required to share their methods so that “simple, clear, sure knowledge is gained” \citep{pierce_whither_1969}.  The system would be formally named the “The Common Task Framework.” Colloquially, it would become known as “benchmarking” (and, derogatorily, as “bake-offs”).

Benchmarking differed from organic evaluation systems used in basic science because it forced the adoption of a single, narrow definition of AI: AI was about creating algorithms that would advance the state of the art on commercial and military problems. In this context, progress was not about resolving theoretical disputes or demonstrating creativity on toy examples. Progress was showing your algorithms could generalize to new datasets and scenarios, as measured by a single, interpretable, quantitative metric: predictive accuracy. This perspective clarified the role of AI scientists, not as philosophers, neuroscientists, or cognitive scientists building holistic general AI, but as engineers working on specific tasks and measuring progress through explicit metrics. 

Benchmarking also pushed the field away from basic science because it required scientists to cede much of the field's autonomy to external actors. In Bourdieu’s view, the independence of scientists to define their own norms/rules, culture, and systems of recruitment is critical to their ability to generate ideas and build bodies of knowledge \citep{bourdieu_rules_1996,bourdieu_science_2004}. But in benchmarking, DARPA chose the problems that were important. DARPA decided how those problems should be broken down into smaller “tasks” that were realistically attainable and measurable. DARPA invited industry labs to compete alongside academics. And DARPA redefined the “game” of science, replacing non-monetary rewards (e.g. publications, awards, and citation) with cold, hard cash. NIST director and white house policy advisor Elham Tabassi explained to us the logic behind this approach: “What causes advancements in technology is when the money start pouring in…that is one way progress happens- when there is one clear application, a clear market, and the challenge has been formulated in a way that it's clear to the community what needs to be done.” 

Mark Liberman, a professor at UPenn and former head of linguistics at Bell Labs, joked that one colleague in ASR complained at the time that formal evaluation was “like being in first grade again.” But at a time when money was scarce, this approach facilitated buy-in.

\subsection{TECHNOLOGICAL ADVANCES IN COMPUTE AND DATA ACCESS}

Freshly transferred from the NSA, enterprising DARPA program manager Charles Wayne saw the potential of Sears’ approach, and organized the first benchmarking competition in 1987 with two algorithms: a symbolic, rule-based AI system from Raj Reddy’s team at Carnegie Mellon, and a statistical learning model called a Hidden Markov Model (HMM) from BBN Technologies. In a surprise to all involved, the HMM edged out the rule-based approach with a slightly higher word translation accuracy. Perhaps it was a fluke. The following year, five more teams were invited and again, the HMM trounced the rule-based system, this time, decisively.

These results were revelatory because they showed that benchmarking could be more than a cudgel used by funders, but a legitimate scientific tool for revealing epistemic insights. Word translation accuracy was acknowledged to be a crude measure that excluded important considerations like a word’s frequency of use, semantics, idioms, or syntactic importance in the sentence. Yet it was able to arbitrate between competing approaches in a way that qualitative debates had failed. Benchmarking showed decisively that statistical algorithms, which could probabilistically handle the ambiguities of real data, were more effective than rule-based ones. 

More generally, it showed the potential of machine learning methods to overcome what Schank saw as a fatal flaw of GOFAI: with more data and computer power, statistical learning algorithms could “scale” in ways that proved impossible for symbolic approaches. HMM’s had been around since the 1960s, but according to John Makhoul (a lead scientist at BBN who helped build HMMs for the 1987-88 competitions), were previously derided by theory-driven linguists as “dumb engineering.” This assessment largely came because computers had not been powerful enough, nor digital data plentiful enough, to train statistical models with so many parameters. 

Over time, benchmarking continued to transform the field. In addition to arbitrating between individual approaches, researchers found the method also effective for charting collective progress in the field over time. “Leaderboards” were established that plotted advances in the state-of-the-art accuracy scores over time. These leaderboards made it simple to distinguish between significant innovations and steady progress. Small gains on accuracy are, in Kuhnian terms, “normal science”—the steady march of progress in constructing collective knowledge. Large leaps in benchmark scores signaled major innovations \citep{kuhn_structure_2012,sim_using_2003}.

Benchmarks first gained traction in the automated speech recognition and natural language processing communities,  but the culture was \textit{contagious}. Competitions quickly spread to other AI communities like information retrieval and computer vision. In 1993 researchers at Bell Labs introduced MNIST, an improved version of a NIST dataset, as a public community benchmark for anyone working on vision problems \citep{bottou_comparison_1994}. MNIST would go on to become one of the most famous dataset in machine learning. Formal benchmarking competitions started to be hosted at computer science conferences without the involvement of NIST or DARPA, sometimes with monetary incentives. Eventually, corporate sponsors began to rival or supplant DARPA/NIST in sponsoring academic conference competitions (e.g., the famous KDD Cup). Over time, state-of-the-art benchmarking results supplanted theoretical novelty as the necessary (albeit, not sufficient) criterion for acceptance at major ML publication venues.

\subsection{CONSTRAINED EPISTEMIC PLURALISM}

It has long been uncontentious in the philosophy of science that contributions are value-laden \citep{mcmullin_values_1982,mcmullin_virtues_2013}. According to \citet{dotan_theory_2021}, “epistemic virtues are theoretical characteristics that are valued because they promote epistemic goals, such as the attainment of truth, knowledge, understanding, or explanation.” From this perspective, the goal of organic evaluation in science is to holistically evaluate contributions across these various epistemic goals. Weighing these different values, researchers provide feedback to improve contributions, and/or bring visibility to those that are deemed most significant. In the GOFAI era, theoretical and material quagmires prevented the field from applying this type of evaluation effectively and developing the types of "doable problems" that enable fractious fields to be productive \citep{fujimura_crafting_1992}.

In the 1990s and 2000s, crowning a single epistemic value (predictive accuracy) over others allowed the new field of machine learning to blossom without forming consensus on the best approach to AI. If an algorithm could achieve state-of-the-art empirical performance, it was worthy of consideration. But the presence of evaluative institutions beyond benchmarking, like peer review and theoretical frameworks, ensured that accuracy was not the only epistemic value that mattered.

\begin{figure}[H]
    \centering
    \includegraphics[width=\linewidth]{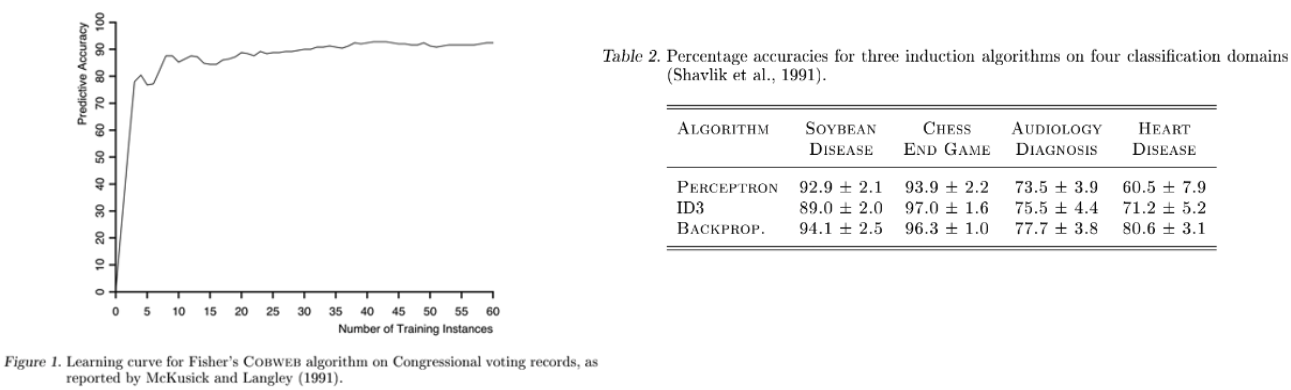}
    \caption{\textbf{Learning Curve (left) and Benchmarking Table (right). }Learning curves, which depict the tradeoff between learning efficiency and accuracy, have been  largely replaced by accuracy-only benchmarking tables in MLR.}
    \label{fig:learningcurve}
\end{figure}

Consider peer review. In his capacity as the first editor of  the field's first journal \textit{The Journal of Machine Learning}, Pat Langley explicitly solicited papers with empirical, theoretical, and psychological evaluation components. He would later write passionately about how different types of figures could be used to assess a machine learning algorithm’s success on different epistemic values. \citep{langley_research_1987,langley_experimental_1997}. For example, one type of plot popular early in machine learning’s history was the “learning curve” (Fig. \ref{fig:learningcurve}). In comparison to a benchmarking table, the learning curve plotted two epistemic values against each other: number of training examples (i.e., data efficiency) and accuracy (i.e., generalization). Expectations of these plots in papers thus incentivized the creation of data-efficient algorithms.

In the second era, unifying theory finally emerged in AI. Instead of building theories of mind, researchers borrowed ideas from statistics to characterize the asymptotic (and idea) behavior of learning algorithms. Interestingly, these frameworks often traded prediction accuracy against other epistemic values. For example, Leslie Valiant, a computational complexity theorist who developed an interest in AI, introduced “Probably Approximately Correct Learning” theory (PAC learning) to the field in 1984 \citep{valiant_theory_1984}. PAC learning allowed scientists to design algorithms that, with high probability, could learn functions with an approximate amount of error. But because of his complexity background, his theory emphasized that learning algorithms should not only be accurate, but also computationally efficient (e.g., run in polynomial time) and data efficient (e.g. require a limited number of training examples).

Another popular theoretical framework, aptly called “statistical learning theory,” (SLT) blossomed after the migration of Soviet mathematician Vladimir Vapnik to Bell Labs’ New Jersey campus in 1990 \citep{vapnik_nature_2000}. Like PAC learning, SLT also described theoretical accuracy bounds of algorithms to solve certain problems.\footnote{called the Vapnik-Chervonenkis dimension in SLT} But it differed from PAC by wedding generalization to a different epistemic value: parsimony. In Vapnik’s view, the generalization ability of an algorithm was a tradeoff with its simplicity (i.e., as measured concretely through the number of parameters or more abstractly through the concept of regularization) \citep{corfield_falsificationism_2009,bargagli_stoffi_simple_2022}.\footnote{Interestingly Vapnik viewed statistical learning theory as addressing the problems of natural science. He frequently compared his emphasis on model regularization to Popper's emphasis on falsificaton. See \citet{corfield_falsificationism_2009}.}

Although predictive accuracy reigned supreme, the diverse values of the field at the time led to the emergence of a variety of machine learning model families with distinct epistemic strengths (e.g., Bayesian networks, Bayesian non-parametrics, decision trees, support vector machines, ensemble models, neural networks). Moreover given the still constrained data and computational resources at the time, each of these approaches was reasonably competitive in terms of accuracy. According to Langley, this made machine learning susceptible to fashions where each family had its moments in the limelight based on some blend of theoretical attractiveness, performance, and hype. Our quantitative analysis corroborates his assessment (Fig \ref{fig:pluralism}). In the 1980s and early 1990s, explainable ``white box" algorithms like decision trees and Bayesian networks were in vogue. In the mid 1990s, support vector machines (SVMs) took the field by storm because they combined empirical performance with attractive statistical properties and guarantees. And in the 2000s, non-parametric Bayesian graphical models emerged as worthy competitors to SVMS by blending all of the above.

A constant competitor throughout this period was a resurgent ``connectionist" neural networks movement led by a core set of collaborators including David Rumelhart, Geoff Hinton, Jay McClelland, Terry Sejnowsky, Yann Lecun, and Yoshua Bengio. Dead for decades after the publication of \textit{Perceptrons}, Rumelhart and McClelland resucitated interest in ANNs with an 1100 page manifesto titled \textit{Parallel Distributed Systems} published in 1987.  The key to their retort was a new wrinkle: hidden layers allowed neural networks to overcome the linear separability problem. Multi-layer networks were attractive on multiple epistemic dimensions. Not only were they human brain-inspired, but connectionists also showed that hidden layers encoded simple ``representations" of their training data \citep{rumelhart_learning_1986}. And with enough neurons, a single hidden layer could theoretically approximate any non-linear function (the fatal flaw of perceptrons) \citep{cybenko_approximation_1989}. The key to these breakthroughs was the (re-)discovery of a new algorithm for training multi-layer networks by Rumelhart called ``backpropagation" \citep{rumelhart_learning_1986}. 

In summary, the diverse theories and methods that emerged during this period laid the groundwork for a new science of machine learning that, even before deep learning, was incredibly productive. These tools are used by millions of data scientists and academicians around the world, even today, in scenarios where training deep neural networks is impractical or other epistemic values are important.

\section{2012 TO 2021: THE ERA OF DEEP LEARNING MONOCULTURE}

In the Era of Symbolic AI, the field puttered as a tribalist protoscience that was unable to overcome epistemic and material constraints to achieve consensus. In the Benchmarking Era, the introduction of formal evaluation through benchmarking organized the field around a narrower vision of  science that allowed for real progress in statistical learning algorithms and theory. Nevertheless, the presence of other evaluative institutions and the material constraints of the time encouraged the creation of a variety of modeling families, each competitive on accuracy but with different epistemic strengths.

\begin{figure}[H]
    \centering
    \includegraphics[width=\linewidth]{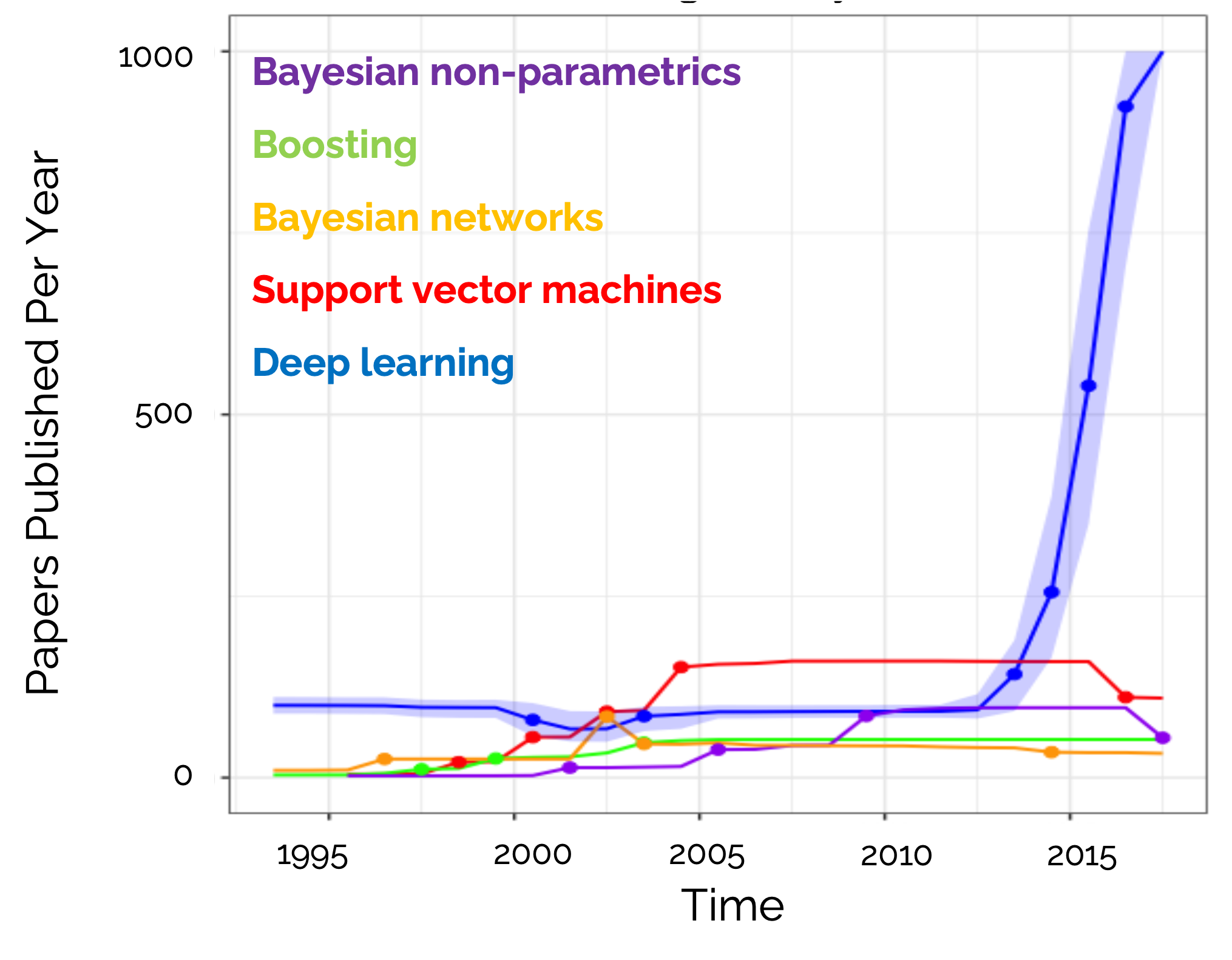}
    \caption{Estimated rates of publication in AI on various different machine learning techniques from 1993 to 2018. Shaded areas represent credible intervals and dots represent statistically significant rate shifts. Deep learning explodes in 2013, while research on other methods remains stagnant.}
    \label{fig:pluralism}
\end{figure}

Everything changed in 2013 (Fig. \ref{fig:pluralism}). Research on neural networks exploded. Histories of AI have largely dramatized this pivot around a single event: the 2012 ImageNet benchmarking competition \citep{gershgorn_dave_data_2017}. In competition with statistical methods, a neural network named AlexNet crushed other types of algorithms with over 10\% lower error than any of the competitors. From a design perspective, AlexNet was not theoretically novel; in fact, it was nearly identical in architecture to Yann LeCun’s 1989 network LeNet. Instead, the model’s key innovation was its ability to be trained on graphics cards. This allowed the authors to make LeNet deeper and wider than ever before: scaling from three hidden layers and 10,000 parameters to eight hidden layers and 60 million parameters. The approach showed convincing evidence that simply scaling up neural networks and using large datasets like Imagenet was a viable method of increasing accuracy on benchmarks.  

Despite parallel trends occurring with other vision datasets and in NLP, the story of the 2012 Imagenet competition is widely viewed as a validation for deep neural networks as a methodology, and the beginning of the “deep learning” era of AI \citep{graves_offline_2008,krizhevsky_learning_2009} . Over the next ten years, AIR would enjoy frenzied progress and explosive growth on the back of deep learning. The driving force behind this growth was a simple reality: the larger the network and dataset, the better a model performed on benchmarks.

In fact, we argue that a distinctive feature of the last decade of AI is not just its explosive progress, but its epistemic narrowness. What we find striking about Figure 3 is not the well-documented explosive growth of machine learning, but instead the fact that the other methodological paradigms introduced at the end of Section 2 effectively faded away after 2013.\footnote{Because scientific production ( especially in AI) grows exponentially, constant levels of research on other model families is an effective decrease in relevance in the field.} This phenomenon can partly be attributed to the conservatisim incentived by benchmarking itself: when a contribution's significance becomes equated with state-of-the-art (SOTA) accuracy, the safest career choice is to demonstrate SOTA results with the least resource investment. In practice, this means starting with the current SOTA (deep learning) and marginally iterating from there.

However, the story is more complicated than that. In this section, we describe how material, organizational, and epistemic factors dovetailed to create a \textit{deep learning monoculture}. From a material perspective, the discovery that reliable performance gains could be achieved by building larger and larger networks created a demand for larger and larger datasets and compute. From an organizational perspective, these greater material needs transferred the locus of power in AI away from government and towards large IT firms (i.e., Facebook and Google) who had the resources, but also less incentive to explore other research strategies.  And from an epistemic perspective, scaling traded off accuracy gains against other epistemic values like explainability, parsimony, and theoretical consistency. The flipside is that this tradeoff relieved reliance on slower evaluative institutions beyond benchmarking (i.e., peer review, formal theory), allowing the field to grow/advance faster than ever before.

\subsection{THE TECHNICAL RAMIFICATIONS OF (REALLY) BIG DATA AND COMPUTE}

The ImageNet 2012 competition was so influential because it revealed two crucial material affordances required to unlock the potential of deep neural networks: parallel processors and big datasets. By the early 2000s, neural networks had fallen out of favor compared to other approaches (particularly SVMs and Bayesian models), largely due to how difficult they were to train (Fig. \ref{fig:pluralism}).  Unlike optimizing other models, training an ANN through backpropagation did not guarantee that the model would learn the best solution from its training data. ANNs thus required art-like fiddling and multiple attempts before acceptable accuracy results could be achieved. Backpropagation on serial computer processors was also slow, yet building custom parallel chips to accelerate the process was impractically expensive. According to Makhoul, these limitations led to reduced interest in connectionism in the late 1990s and early 2000s as SVMs and Bayesian non-parametrics became popular (Fig. \ref{fig:pluralism}).

The first breakthrough to these problems came in 2007 when Andrew Ng and students demonstrated that the training of neural networks could be vastly accelerated by using NVIDIA's graphical processing units (GPUs), originally designed for video games and movie effects \citep{raina_large-scale_2009}. The second secret ingredient was the rise of large datasets in computer vision like CIFAR-10, and of course, Imagenet \cite{deng_imagenet_2009,krizhevsky_learning_2009}. At the time, ImageNet consisted of 1.3 million images, each labeled with the objects inside of them. A dataset of this scope was completely unprecedented in the history of academic machine learning research because it had been prohibitively expensive and time consuming to label that many images.\footnote{In 2008, Imagenet became one of the first major projects to crowd-source data annotation, by leveraging the new Mechanical Turk platform. Mechanical Turk allowed researchers to pay below minimum wage for data annotation. Researchers have expressed regret over this point, but the reality is that project otherwise could not have been done in academia \citep{gershgorn_dave_data_2017} .}

AlexNet was not the first neural network to leverage these advantages to build deeper neural networks, but it was the first to show that doing so could yield a network capable of annihilating other machine learning approaches on a high-profile benchmark competition. Other model families were simply not able to exploit graphical processing units or larger datasets to the same extent. 

This discovery that wider and deeper networks performed better on benchmarks, colloquially called “scaling,” has been a \textit{major} driving factor behind the incredible progress of deep learning over the past ten years. In 2022, Andrej Karpathy (cofounder of OpenAI, Director of AI at Tesla) revisited LeNet and AlexNet to celebrate the 33rd birthday of  the former algorithm \citep{andrej_karpathy_deep_2022}. Despite some tweaks, his overall conclusion was,

\begin{quote}
not much has changed in 33 years on the macro level. We’re still setting up differentiable neural net architectures made of layers of neurons and optimizing them end-to-end with backpropagation and stochastic gradient descent. Everything reads remarkably familiar, except it is smaller.
\end{quote}

Between 2012 and 2017, researchers did engage in some architectural exploration (e.g., \citet{cho_learning_2014,he_deep_2015,sabour_dynamic_2017}). But two other transformative innovations propelled scaling forward and put the search for new models to bed. The first was the discovery  by \citet{mikolov_distributed_2013} at Google that deep learning models could learn meaningful correlations in data by using mindless objectives like predicting the next word in all of Google Books. Because this process did not require laborious and expensive data labeling, it unlocked the usage of far larger datasets than even Imagenet. These ``pretrained" models could be ``finetuned" for specific tasks or ``aligned" to human language/values later.

The second breakthrough was the “Transformer” architecture by  \citet{vaswani_attention_2017} (also at Google) that leveraged the structure of text, image, and graph data to greatly improve ANN performance. Where previous designs (i.e., recurrent neural networks) had treated text only as sequences of words, Transformers could attend to relationships between each word in a passage heterogeneously and in any order. However these advantages again came at the cost of other epistemic values. Where the memory and computational complexity of  recurrent networks scaled \textit{linearly} with the length of input text, the material needs of transformers scaled \textit{quadratically} due to a higher density of parameters \citep{gu_mamba_2023}. The higher density of parameters in Transformers also made it harder to explain what, or understand how, these learned from data. \textit{After 2017 and before 2021, there were essentially no major innovations in the design of neural networks.} To understand why, we turn to the organizational forces that drove the field toward scaling.

\subsection{ORGANIZATIONAL POWER SHIFTS FROM ACADEMIA AND GOVERNMENT TO INDUSTRY}

The new emphasis on data and compute created by scaling shifted the locus of power within the field. Over the 1990s and 2000s, the growing utility of machine learning methods had started to make technology companies increasingly important actors within the field. While government funders were still key players, companies like Yelp and Netflix began to fund benchmarking competitions \citep{yelp_yelp_nodate,netflix_netflix_2009}. For young academics, both industry and academia were seen as viable career paths.

As data and compute became increasingly important, two companies were uniquely positioned to capitalize on machine learning and dominate the field: Google and Facebook (now Meta).  Data is the foundation of Google and Meta’s business. Innovated by tech executive Sheryl Sandberg first at Google and later at Facebook, both companies share the same “surveillance capitalism” revenue model: they provide free services (e.g., Google Search, Gmail, Android, Facebook, Instagram) in exchange for customers’ personal data \citep{zuboff_big_2015}. This data is used to develop targeted advertising platforms to sell to other companies. Because both companies have billions of customers, their troves of data were/are enormous and their computing infrastructure peerless. Moreover, they had a vested interest in machine learning to create both compelling services and effective advertising. 

Seeing the potential of deep learning, Google, Facebook, and others invested heavily in industrial AI research labs. Google founded Google Brain in 2011, and poached Andrew Ng and Geoff Hinton from academia to run it. It later bought Deepmind in 2014. Facebook hired Yann LeCun to start Facebook AI Research in 2013. Realizing Microsoft Research was behind the curve, the company invested billions of dollars in startups OpenAI and Cohere between 2019 and 2023. In China, large IT firms such as Alibaba and Baidu have also started major machine learning labs.

The large investments of these IT firms significantly reduced the influence of government in driving the research agenda of the field. As Ken Church, former director of NLP research at Baidu said, “NIST no longer matters.” In 2022, the US government spent more than 6 billion dollars on AI research. Meanwhile, the tech giants Google, Microsoft, and Meta spent at least 20 billion alone.

It is not the case, however, that industry made academia irrelevant. It became the norm for industry to publish academic papers and open-source their benchmark datasets and algorithms. This practice kept academics involved in researching questions that were of direct interest to companies, and maintains academia’s relevance as a PhD training pipeline. Ken Harper, a Vice President at Nuance (a Microsoft subsidiary), explained, "Everyone's using the same technology from a modeling perspective. What's different is the data and how you teach these machines with the data." Yoshua Bengio has expounded that industry benefits from this arrangement because the value during the scaling years was not in the models, which rarely changed. It was in the large, proprietary datasets used to train them \citep{yoshua_bengio_mind_2017}.

Nevertheless, the concentration of data, computing, talent, and money at large IT companies has firmly shifted the power to set the agenda of the field into the hands of industry, not academia or government. In other work, we have shown how industry and elite academic affiliates produce the majority of benchmarks that are used across the field \citep{koch_reduced_2021}. To the extent that benchmarks define the tasks and data that the field should seek to solve, this is a reflection of industry’s disproportionate influence on the field. Others have shown that industry researchers consistently produce the top benchmarks \cite{ahmed_growing_2023}.

Multiple interviewees at Apple, Google and Microsoft have emphasized how the disproportionate influence of large industry players has ossified the field around a scaling monoculture. It is no secret that large organizations, government or corporate, are conservative in their agendas. In the context of science, Naomi Oreskes has illustrated how large organizations are naturally inclined to support research trajectories that align with their interests \citep{oreskes_science_2021}. For corporations, exploring high-risk and/or basic research is not good business. On the other hand, scaling was, and remains, the intellectually cheapest way for companies to create more accurate algorithms.  As a Principal Scientist at Apple who asked to remain anonymous put it,

\begin{quote}
The path of least resistance in terms of a human cognitive breakthrough is to keep using the industry approach of adding more data and computation… Maybe the improvement [on benchmark scores] will diminish, but at least you are guaranteed. Whereas compared to really having the intellectual breakthrough, that's really unpredictable.”
\end{quote}

Francois Chollet, one of the primary inventors of Google’s Tensorflow (the most widely-used deep learning programming language in the world) agreed:

\begin{quote}
I think we're going to develop new directions again when we've run out of steam with the current approach. The only issue is that of course you can always squeeze some extra juice out of scale. Like with a few extra billion dollars you can probably get something extra. And so far there's been this insane appetite to just scale more, throw more money at the problem, and of course it is a diminishing return. I mean, we've been in diminishing returns territory for years now. So we've had to move from \$1 million training runs to \$100 million training runs. And the models aren't 100x better. They're maybe 2X better. But yeah, what about 1 billion dollar training runs? What about 10 billion training runs? Maybe they can be 20\% better and apparently we are still going because there's still appetite for that. But at some point we will have to explore new things.
\end{quote}

\subsection{THEORY AND PEER REVIEW IN A BENCHMARKING WORLD}

While only evident in retrospect, the commitment to scaling represented a unique epistemic bargain: the field exchanged singular accuracy gains on relevant tasks for other epistemic values. Consider the values previously wedded to accuracy in the field's PAC learning and SLT theoretical frameworks. Both theories emphasized identifying theoretical bounds on accuracy in the limit, but ANN optimization through backpropagation makes no convergence guarantees, rendering such discussions somewhat irrelevant. In practice, training large ANNs requires tacit knowledge, intuition, trial-and-error, and potentially thousands of false starts to get a network to converge on an acceptable solution. PAC learning emphasized efficiency in terms of both data and compute. In contrast, scaling relies on thousands of GPUs gobbling up electricity and internet-scale datasets. A key value in SLT is parsimony. But as ML theorist Rishi Sonthali told us, ``in classical [low dimension] statistics... we want the correct sort of complexity level and that’s the only thing that will work. But modern machine learning is like nah, just make it bigger. And somehow nothing bad happened." In other words, enormous networks are actually better at prediction than smaller ones.\footnote{This phenomenon is called ``double descent" in the machine learning literature. For interested readers, see \citet{belkin_reconciling_2019}.} Lastly, simpler models are preferred in science when explanatory power is similar (i.e., Occam’s razor), in part because they are more interpretable. In practice, networks with billions of neurons have limited explanatory power, and they are completely opaque to mechanical interpretation.\footnote{The field of ``explainable AI" has made little progress on mechanical interpretation of parameters. Some of the most commonly used methods in the field like SHAP and integrated gradients have been criticized as misleading. See \citet{bilodeau_impossibility_2024}.}

At the awards ceremony for NeurIPS (one of most prestigious AIR publication venues) in 2017, Ali Rahimi and Ben Recht were honored with a “test-of-time award” for their work on efficiently training SVMs \citep{rahimi_ali_reflections_2017,rahimi_random_2007}. In their speech, they lamented how the “NeurIPS rigor police” had kept them honest about the asymptotic bounds of their algorithm in 2007. In contrast, they compared the current focus on deep learning to “alchemy,” with a special emphasis on how training deep networks has become more art than science because our understanding of their learning and decision-making is so limited.

The critique of deep learning as alchemy, while inflammatory at the time, is now widely acknowledged by elites within the field. In the conclusion of an influential 2021 paper, Noam Shazeer, an inventor of the Transformer, cheekily concludes, “we offer no explanation as to why these architectures seem to work; we attribute their success, as all else, to divine benevolence” \citep{shazeer_glu_2020}. Prominent AI critic Melanie Mitchell asked us, “Modern deep learning, does that have a theory? I don’t think it does […] It seems like trial and error.” And to quote Chollet,
\begin{quote}
We build systems, we don't do anything else. We don't really have theory. We build systems, and then these systems are defined by what they can do, and they're better than other systems if they can do more. And you have to use benchmark data sets to test that. So that's really the first thing to understand is that we are not, we're not doing science here. We're doing engineering.
\end{quote}
Yann LeCun was livid with Rahimi and Recht's speech, but \textit{not} at the claim that AI has become alchemy. Instead, he questioned the utility of theory altogether, 
\begin{quote}
Sticking to a set of methods just because you can do theory about it, while ignoring a set of methods that empirically work better just because you don't (yet) understand them theoretically is akin to looking for your lost car keys under the street light knowing you lost them someplace else.
\end{quote} 

There is, of course, still theory driving design in AI, but it is intuitive theory. While moving away from mathematical formalization may be a shift for AIR, it is not exceptional in science by itself. But LeCun's critique hits on a key strength of epistemic monoculture: \textit{when there is only one epistemic value consistent with progress, organic evaluation institutions that bridge multiple values (e.g., formal theory, peer review) are optional. All you need is benchmarking.} Beyond theory, the relevance of peer review has been diminishing in machine learning. As of 2021, NeurIPS expects papers to cite relevant pre-prints released within three months of the submission deadline. And prestige corporate AI labs like OpenAI submit their white papers for publication only occasionally. Organizationally, formal theory-building and peer review are on the decline because they are slow, laborious processes. On the other hand, the speed, immediate verifiability, and easy intelligibility of benchmarking have been key factors in the field's unprecedented acceleration over the past decade.   

\section{DISCUSSION}

In the preceding pages, we chart the historical development of the field of AIR through three eras and across their epistemic, organizational, and technological dimensions. In the Era of Symbolic AI, theories of machine intelligence flourished, but technological limitations and misaligned organizational incentives produced a fractured field with little way for organic evaluation to take hold. During the Benchmarking Era, enterprising DARPA program managers redesigned the field around formal evaluations based on narrowly defined tasks with commercial or military application. Benchmarks highlighted the effectiveness of statistical machine learning approaches over the symbolic ones that dominated the field at the time. For the next 25 years, several competitive machine learning model families emerged, each manifesting different epistemic strengths and weaknesses. However, advances in computer technology and increased access to data through the internet produced the ideal conditions for one approach- deep learning- to dominate benchmarks.  The Deep Learning Era was characterized by three significant developments. First, a narrowing of the field to deep learning led to a deep commitment to the scaling of compute and data that allowed it to make progress. Second, there was a significant shift of gravity away from government and academia to the tech industry, who had the data and compute to pursue scaling as a research program. Third, the reliance on benchmarking diminished emphasis on organic evaluation institutions such as formal/mathematical theory-building and peer-review.

Put even more simply, intellectual fecundity and chaos of the first era motivated the creation of benchmarks, a muscular organizational intervention to structure the field. Benchmarks, in turn, provided the perfect soil for deep learning to out-compete and dominate a landscape shaped by formal evaluation.

This historical narrative illustrates how the organizational structure, epistemic orientation, and material resources of science can dovetail to ossify an epistemic culture that is simultaneously extremely efficient and extremely narrow in its research interests (i.e., a monoculture). In AIR, the ceding of autonomy and the adoption of benchmarking were the critical connective tissues in this alignment. Ceding autonomy allowed external actors to streamline research agendas, and benchmarking was the technological intervention that allowed them to do so. 

A monoculture like AIR is efficient because benchmarking directs researchers to optimize a single epistemic value on specific tasks, and provides simple, verifiable, and interpretable quantifications of significance/progress towards those goals. The slow process of self-directed exploration in basic science can be accelerated because the externally-dictated problem space for scientists to explore is smaller.\footnote{Yann LeCun has made a similar point.}  Plodding organic evaluation institutions like peer review, mathematical theory-building, and citation can be shed because there is no emphasis on innovating across multiple epistemic values.

But tying its identity to a single epistemic value and external interests is also how AIR became so narrow in its research agenda. When scientists can only gain high-status publications by demonstrating SOTA accuracy, the safest research choice becomes incrementally advancing proven methods, not innovating new ones \citep{foster_tradition_2015}. In basic science, organic evaluative institutions allow fields to evolve across different epistemic dimensions. Furthermore, external actors like government and industry are also beholden to their own stakeholders. Industry in particular must pursue research programs that will yield immediate applications at the lowest financial risk, regardless of their long-term potential. To this point, recent research has suggested industry R\&D is less innovative than work by academics in AIR \citep{liang_complementary_2024}.

The final ingredient that ossified this arrangement as a monoculture and not a transitive state of the field is the distribution of material resources. Resources are power. The emergence of scaling as the field's favored research program wrested power away from government and academia because they lacked the data, compute, or money to participate autonomously. Academia has subsequently become a training pipeline for industry, and academic research is dependent on industry for grants, data, pre-trained models, and access to compute \citep{ahmed_growing_2023, koch_reduced_2021}.

\subsection{THE SOCIAL IMPLICATIONS OF AIR MONOCULTURE}

AIR's monoculture is sociologically provocative because it poses a challenge to the 20th-century belief that basic research is necessary for scientific progress. Whether monocultures are normatively better or worse than basic science at knowledge construction is unanswerable. Nevertheless the empirical outcome of AIR's monoculture, large-scale deep learning, has significant social impacts, both positive and negative.

Given the emphasis on accuracy above all else, it is unsurprising that large-scale deep learning rates weaker on many of the epistemic values that are traditionally prized in science \citep{mcmullin_values_1982,mcmullin_virtues_2013,dotan_theory_2021}.   Because scaling progresses by fueling ever-more parameters through ever-more compute and data, the approach is inherently less parsimonious, explainable, or data/compute efficient than others popular ten years ago. From a realism perspective, text pre-training is decidedly alien from human learning, requiring millions of repeated exposures to millions of data points, instead of a handful of exposures to a handful of data points. To elide these limitations, the field has moved away from mathematically formal (a la physics or statistics) and/or realist (a la neuroscience) theory-building  and relied more on empirically-testable, intuitive theory to drive innovation. 

Real negative social consequences result from these epistemic weaknesses.  First, the carbon footprint of AI is enormous. A recent estimate suggested AI will soon consume as much electricity as Argentina or Sweden \citep{vries_growing_2023}. It is not just training that contributes to this cost, but inference: generating a single image from a text prompt consumes about as much electricity as charging a cellphone to 100\% \citep{luccioni_power_2023}. Second, the curation of large datasets often requires tacit privacy violations when IT companies use surveillance capitalism to harvest data from social media, apps, search tools, and devices  \citep{zuboff_big_2015}.  Another important source of data, webscraping, has similarly spurred lawsuits by content creators seeking royalties for their intellectual property \citep{small_sarah_2023}. Finally, the inability to understand how these models learn or make decisions has perpetuated social inequalities. Six years after AI ethicists first called attention to how language models gender women as nurses rather than doctors, generative AI models prompted to create images of "terrorists" and "criminals" still almost exclusively synthesize pictures of Brown men \citep{caliskan_semantics_2017,nicoletti_humans_2023}.  If AI required smaller, more-tractable datasets or made explainable decisions, we could better mitigate some of these deleterious consequences.

However, AI ethics concerns have (regrettably) begun to fade into the limelight as the accomplishments of large-scale deep learning pile up. The deep learning era already saw the creation of classifiers that outperform experts at limited tasks (e.g., detecting tumors in medical imaging) and social media/search algorithms that crucially shape how we consume cultural content and knowledge. Now we are entering a new era of generative AI. In the past two years AI has moved beyond creating human-level cultural content to independently solving open-ended, high-skill tasks (e.g., taking doctors' notes, designing websites). Perhaps the greatest argument for monoculture then is that this approach \textit{is} moving us closer to the field's original theoretical dream of general AI. This is happening not because scientists were allowed to explore autonomously, but because of their focus on commercial tasks. It turned out that passing the Turing test required a market for artificial customer service agents, not a theory of a mind.

\subsection{THE DIFFUSION OF THE DEEP LEARNING MONOCULTURE ACROSS THE SCIENCES}

So what does the success of monoculture mean for the future of basic research in science? Basic research is obviously not going away. But our contention is that AI's monoculture is \textit{contagious}.  Over the past twenty years, sciences have already begun to change their workflows to account for the size and ``black-box" nature of machine learning algorithms. But compared to older models, deep learning and generative AI require even greater compromises along these dimensions. We believe these compromises will further spread some of the epistemological and organizational features of AIR's monoculture to other fields.  

Epistemologically, black-box models have already spurred debates across the sciences about the values of prediction \textit{vis à vis} explanation. In statistics, Leo Breiman's piece about ``the two [evidential] cultures" has over 5,700 citations \citep{breiman_statistical_2001}. In sociology and computational social science, Duncan Watts and collaborators have similarly pushed for greater emphasis on generalizability over “unscientific stories” (2014:313) \citep{hoffman_managing_2017,watts_common_2014}.  Machine learning models have typically been applied for intermediate problems that are deemed taskifiable, and/or where the benefits of prediction outweigh the utility of explanation. But as deep learning and generative AI progress to increasingly sophisticated tasks, decisions about which problems can be taskified and which ones demand theoretical explanation will become harder. 

The story of AlphaFold, deep learning's crowning scientific success  in science to date, illustrates this dilemma. AlphaFold is a Google Deepmind algorithm for predicting the 3D structure of proteins. Folding proteins was considered a famously difficult problem in Biology. Capturing the crystal structure of a protein \textit{in vivo} is a slow and laborious process, and biophysical models had not achieved the maturity to reliably simulate structures across a wide variety of contexts.

As structural biologist Mohammed Al Quirishi writes, protein folding was perfect for DeepMind because it was already taskified "with clear objectives and metrics. Science is almost never this way but protein structure prediction actually fit the bill perfectly." Every other year, structural biologists would hold a benchmarking competition to compare their simulations against crystalized structures for proteins \citep{alquraishi_alphafold2_2020}. Google entered the competition twice, and in 2021, effectively solved the protein-folding task. 

The success of AlphaFold has transformative potential in fields as varied as neuroscience, evolutionary biology, and drug discovery. But the reality is that while the “task” of protein folding has been solved, we are no closer to understanding how folding occurs than before AlphaFold’s success \citep{chen_opinion_2023}. Structural biology thus faces a crossroads that other sciences will increasingly have to confront: do they build on AlphaFold’s opaque solution and move onto other problems, or continue to work on protein folding as a pressing theoretical issue? As black-box deep learning is applied to increasingly sophisticated problems in science, the debate between predictive and explanatory evidential cultures will become ever-more urgent \citep{collins_meaning_1998}.

Because unexplainable AI can only be evaluated through benchmarking, this epistemic dilemma also has implications for autonomy and the organization of science as well. Can benchmarking competitions bring progress to other problems where explanation-based, basic science has failed to build consensus? In 2017, \citet{salganik_measuring_2020} organized “the Fragile Families Competition” as a radical test of this approach in the social sciences. It turned out that machine learning models of the time were not able to overcome the sparsity of that dataset. But outcomes of such experiments may change as models evolve in the era of Generative AI. For a field like sociology that has achieved limited consensus on how to explain social phenomena, more benchmarking might be a good thing.  Sociologists might chafe at the loss of autonomy if publication in \textit{American Sociological Review} required achieving the highest R\textsuperscript{2} on an inequality task. But it might help the field elide the intractability of its project.

The data and compute requirements of scaling are also driving organizational changes in science. Data has traditionally been viewed as private, high value, and the source of competitive advantage in science. But the rise of large datasets (and large models that require them) has pushed traditionally low-data fields towards centralized repositories similar to those that emerged in genomics thirty years ago. At the same time, scientific reform movements have pressured publishers and scientists to publicly contribute their data to these repositories to address the replication crisis \citep{peterson_metascience_2023}.

As in AI, steep data and compute requirements are also increasing the dependence of all fields on the IT industry.\footnote{Not to mention social scientists' increasing reliance on social media data as our social interactions have moved online.} In the deep learning era, scientists started to become dependent on models that were pre-trained by Facebook and Google on proprietary datasets (e.g., Word2Vec, BERT). When these models need further fine-tuning, scientists often turn to Amazon and Microsoft for cloud compute resources. And in the era of generative AI, the sheer scale of models has led IT companies to transition from public releases to subscriptions and a la carte fees for use.\footnote{Open models are discussed in the conclusion.} We are reaching the point where universities must cede autonomy with respect to compute or be left behind. This has led to recent initiatives at the national and international levels for public-funded supercomputers for LLM training and inference that are in their early stages \citep{lohr_universities_2020,workshop_bloom_2023}.

We note that the processes of contagion described here go far beyond science. Technology-heavy fields like finance and medicine are struggling with the same epistemic and organizational dilemmas stemming from their increased reliance on black boxes and the tech industry. As Generative AI evolves, similar challenges will be presented to each of us in our daily lives as well.

\subsection{CONCLUSION: THE TRADEWINDS OF GENERATIVE AI}

 Our history ends in 2021 because AI is presently being transformed by generative models. As we reflect above, the success of generative AI challenges basic science by showing that monoculture can spawn algorithms that are not only narrowly task-driven, but move us closer to the field's original goals of general AI.  It is too early to tell, but several of the evaluative and organizational features of AIR's monoculture may be in flux due to this disruption.

First, carefully constructed, mono-value benchmarks may not be able to keep pace with the expanding capabilities of large models. One popular alternative right now are dynamic, crowd-sourced ``chess match" benchmarks where users enter any prompt they like and pick the better response from two large language models \citep{chiang_chatbot_2024}. A chess ranking algorithm then sorts models on these wins and losses to create leaderboards. Like citation or peer review, this new form of benchmarking is organic in that users are likely considering multiple epistemic values. But the values they are considering, or their expertise in making those judgements, is unclear, turning benchmarking into a black-box itself. At the same time, large models have created a crisis for laborious organic evaluation in science. Recent estimates indicate that up to 15\% of peer reviews in top AI conferences contain some LLM generated content, suggesting that the practice may need to be radically reconsidered \citep{liang_monitoring_2024}.

In 2022, the loss of autonomy by academic AI hit a nadir. After 2020's GPT-3, OpenAI (the leader in the field at the time) stopped releasing full papers and code for it's models to academics as it started to commercialize its models. Academia's role in SOTA research was increasingly murky. But the punishing expense of scaling has also spurred real innovations in model design, training, and inference that can squeeze more performance from the same compute and data.\footnote{Examples include quantization}  In February 2023, Meta broke with OpenAI (and followed past tradition) to release Llama to the public. Llama leveraged these efficiency developments to become the first near-SOTA model that could be run on regular computers for years. Meta has positioned Llama and similar models as ``open" in the sense that they use non-proprietary data and the weights and architecture are free to use. The rise of open models has given academic research much-needed oxygen to help industry move generative AI forward. But it has not increased academic autonomy in any fundamental sense. It is estimated that training Llama 2's 70 billion parameters from scratch costs between 1-2 million dollars, putting training far beyond the reach of academics \citep{touvron_llama_2023}. And as long as scaling remains important, open models will always lag behind the true, enterprise-scale SOTA algorithms (e.g., GPT-4, Gemini Pro as we write this).

It is too early to tell how the rise of generative AI will reshape AIR, and science more generally. We, like all others, are eager to see what the future holds.

\bibliographystyle{asr}
\bibliography{main}
\end{document}